\documentstyle[preprint,aps]{revtex}
\begin{document}
\title{Gauge Fields and Pairing in Double-Layer Composite Fermion Metals}
\author{N.~E.~Bonesteel}
\address{
National High Magnetic Field Laboratory and Department of Physics, 
Florida State University,
Tallahassee, FL 32306-4005}

\author{I. A. McDonald}
\address{Department of Physics, Pennsylvania State University,
University Park, PA 16802}
\author{C. Nayak}
\address{Department of Physics, Joseph Henry Laboratories, Princeton
University, Princeton, N.J. 08544}

\maketitle

\begin{abstract}
A symmetrically doped double layer electron system with total filling
fraction $\nu=1/m$ decouples into two even denominator ($\nu=1/2 m$)
composite fermion `metals' when the layer spacing is large.
Out-of-phase fluctuations of the statistical gauge fields in this
system mediate a singular attractive pairing interaction between
composite fermions in different layers.  A strong-coupling analysis
shows that for any layer spacing $d$ this pairing interaction leads to
the formation of a paired quantum Hall state with a zero-temperature
gap $\Delta(0) \propto 1/d^2$. The less singular in-phase gauge
fluctuations suppress the size of the zero-temperature gap, $\Delta(0)
\propto 1/\left({d^2}{(\ln d)^6}\right)$, but do not eliminate the
instability.
\end{abstract}


Composite fermions were introduced by Jain in order to understand the
observed hierarchy of states in the fractional quantum Hall effect
(FQHE) \cite{jain}. A composite fermion is an electron confined to
move in two dimensions and tied to an even number of statistical flux
quanta.  Jain showed that the fractional quantum Hall effect for
electrons at odd-denominator filling fractions can be viewed as an
effective {\it integer} quantum Hall effect for composite fermions.
Halperin, Lee, and Read (HLR) took Jain's suggestion further, arguing
that at Landau level filling fraction $\nu=\frac{1}{2}$, or any even
denominator filling fraction $\nu=\frac{1}{2m}$, the statistical flux
attached to composite fermions can, at the Hartree level, exactly
cancel the physical flux of the applied magnetic field
\cite{hlr}.  The composite fermions then form a new type of metal, and
a growing number of experiments appear to support this description
\cite{1layer_experiment}. HLR also showed that fluctuations of the
statistical gauge field in this metal give rise to singular inelastic
scattering of sufficient strength to lead to a breakdown of Landau
Fermi liquid theory.  Though experimental proof of the non-Fermi
liquid nature of the composite fermion metal remains elusive, it has
generated a great deal of excitement in the theoretical community
\cite{1layer_theory}.

Double-layer electron systems have been realized in both double
quantum wells \cite{murphy} and wide single quantum wells \cite{wsqw}.
The $(m,m,n)$ states at filling fraction $\nu=\frac{2}{m+n}$, proposed
by Halperin \cite{bih}, are double-layer generalizations of the
Laughlin states. At even denominators, there are additional
possibilities motivated by the composite fermion construction. As one
of us has pointed out \cite{neb}, in the limit where the layer spacing
$d$ is large it should be possible to view a double-layer system at
$\nu=\frac{2}{2m}$ as two decoupled $\nu=\frac{1}{2m}$ composite
fermion metals.  This description will be referred to as the
double-layer composite fermion metal (DLCFM) description in what
follows. The main result of this Letter is that an {\it ideal} DLCFM,
by which we mean a DLCFM in which the carrier densities in the two
layers are precisely equal, there is no interlayer tunneling and there
is no disorder, is {\it always} unstable to the formation of a paired
quantum Hall state for any layer spacing $d$.  Motivated by this
result we propose the phase diagram shown in Fig.\ (1) for the $\nu=1$
double-layer system.  It is unclear at present whether the line
separating the paired quantum Hall state and the $(1,1,1)$ state
represents a transition or a smooth crossover.

The Lagrangian density for an ideal DLCFM as defined above is given by
\begin{equation}
{\cal L} ({\bf r},\tau) = 
{\cal L}_1 ({\bf r},\tau) + {\cal L}_2 ({\bf r},\tau)
\label{lagrangian}
\end{equation}
where
\begin{equation}
{\cal L}_1 ({\bf r},\tau) =
\sum_s \Biggl( \psi^{\dagger}_{(s)} ({\bf r},\tau)  
\bigl(\partial_{\tau} + i a_0^{(s)}({\bf r},\tau)\bigr)
\psi_{(s)}({\bf r},\tau)  
+ {1 \over 2 m_b} \psi^{\dagger}_{(s)} ({\bf r},\tau)
\bigl(-i \nabla -
{\bf a}^{(s)}({\bf r},\tau)  
+ e {\bf A}({\bf r})\bigr)^2 \psi_{(s)}({\bf r},\tau)\Biggr)
\end{equation}
and
\begin{equation}
{\cal L}_2 ({\bf r},\tau)=
-\sum_{s,s^{\prime}} {i \over 2 \pi }\,{K^{-1}_{s s^{\prime}}}\,
a_0^{(s)} ({\bf r},\tau) \epsilon_{ij} \partial_i a_j^{(s^{\prime})}
({\bf r},\tau) 
+ {1 \over 2} \sum_{s,s^{\prime}} \int d^2 r^{\prime}
\delta \rho_{(s)} ({\bf r},\tau) V_{s,s^{\prime}} ({\bf r}-{\bf r}^{\prime})
\delta \rho_{(s^{\prime})} ({\bf r}^{\prime},\tau).
\end{equation}
Here $s$ is a layer index, $V_{s,s^\prime}(\vec r) = e^2/
\epsilon \sqrt{\vec r^2 + (1-\delta_{s,s^\prime})d^2}$ is the intra- ($s=s^\prime$) 
and inter- ($s\ne s^\prime$) layer Coulomb interaction, $\psi_s$ is
the fermion field in layer $s$ and
\begin{equation}
\delta \rho_{(s)} ({\bf r},\tau) 
= \psi^{\dagger}_{(s)}({\bf r},\tau) 
\psi_{(s)}({\bf r},\tau)-n_{(s)}
\end{equation}
is the density fluctuation about the mean density $n_{(s)}$ in layer
$s$.  We work in the transverse gauge, $\nabla \cdot {\bf a}^{(s)}
({\bf r},\tau)=0$, and take $K_{11} = K_{22} = 2m$, $K_{12} = K_{21} =
0$, which is the natural choice in the limit of large layer spacing.
We further specialize to $m=1$, but all results may easily be
generalized.  Integrating out the $a_0^{(s)}$ fields enforces the
constraint
\begin{equation}
\frac{1}{2\pi}\,\nabla \times {\bf a}^{(s)}\, =\, 2\, \delta \rho_{(s)}
\end{equation}
which attaches two flux tubes of the appropriate statistical flux to
each electron.  In the following, we shall denote by $a^{(s)}$ the
fluctuation in the {\it transverse} gauge field associated with layer
$s$.

It is natural to describe the fluctuations of this system in terms of
in-phase and out-of-phase modes.  If the in-phase and out-of-phase
gauge fields are defined as $a^{(\pm)} = a^{(1)}\pm a^{(2)}$, then
within the random-phase approximation the relevant gauge field
propagators at low frequency and long wavelengths are, in the limit $d
\gg l_0$,
\begin{equation}
D^{(+)}(q,i\omega_n)\simeq \Bigl(e^2 q/4\pi\epsilon +|\omega_n|
k_f/(4\pi q)\Bigr)^{-1}
\label{inphaseprop}
\end{equation}
for the in-phase gauge fluctuations, and
\begin{equation}
D^{(-)}(q,i\omega_n) \simeq 
\left\{
\begin{array}{cc}
\Bigl(e^2d q^2/4\pi\epsilon  + |\omega_n|k_f/(4\pi q)\Bigr)^{-1} & 
\mbox{for $q \alt d^{-1}$} \\
\Bigl(e^2q/4\pi\epsilon +|\omega_n|k_f/(4\pi q) \Bigr)^{-1}    & 
\mbox{for $q \agt d^{-1}$}
\end{array}
\right.
\end{equation}
for the out-of-phase fluctuations \cite{neb}.  The current-current
interactions mediated by these gauge fields in the interlayer Cooper
channel are then
\begin{equation}
V^{\pm}_{12}(k,k^\prime;i\omega_n) =
\pm
\left(\frac{{\bf k}\times{\bf \hat q}}{m^*}\right)^2
D^{\pm}(q,i\omega_n)\label{veff}
\end{equation}
where ${\bf \hat q} = ({\bf k} - {\bf k^\prime})/|{\bf k} - {\bf
k^\prime}|$ and $m^*$ is the effective mass of the composite fermions.
Fluctuations in $a^{(-)}$ are more singular at low frequencies because
the Coulomb interaction suppresses the in-phase density fluctuations
but not the out-of-phase density fluctuations.  As a consequence the
effective interaction is dominated by the out-of-phase fluctuations.
We will include both the in-phase and out-of-phase fluctuations in our
calculations, while ignoring the less singular density-density and
density-current interactions \cite{neb}.

The dominant out-of-phase mode mediates an {\it attractive} pairing
interaction between composite fermions in opposite layers.  This
attractive pairing interaction appears because $a^{(-)}$ couples to
composite fermions in different layers as if they were oppositely
charged.  The fluctuating $a^{(-)}$ field strongly inhibits the
coherent propagation of a single composite fermion while a pair made
up of composite fermions from different layers is neutral with respect
to $a^{(-)}$.  Such a composite fermion pair can then propagate
coherently through the fluctuating $a^{(-)}$ field, much like a meson
propagating coherently through a strongly fluctuating gluon field.

In \cite{neb} it was proposed that this attractive interaction might
lead to a `superconducting' instability of an ideal DLCFM.  Such a
`superconducting' state of composite fermions would be incompressible
and thus exhibit the FQHE \cite{zhang}.  Here we investigate this
possibility within the framework of Eliashberg theory.  Using the
Nambu formalism the matrix Green's function is
\begin{equation}
G(k,i\omega_n) = [i\omega_nZ_n-\epsilon_k\tau_3-\phi_n\tau_1]^{-1}
\end{equation}
where $\omega_n = (2n+1)\pi T$ is a fermion Matsubara frequency, $Z_n$
is the mass renormalization, $\phi_n$ is the anomalous self energy,
and $\Delta_n = \phi_n/Z_n$ is the gap function.  The Eliashberg
equations for $l$-wave pairing in this system are then given by
\begin{eqnarray}
\omega_n(1-Z_n) &=& -\pi T\sum_m \frac{Z_m\omega_m}
{(|Z_m\omega_m|^2+\phi_m^2)^{1/2}} (\lambda^{(+)}_{m-n,0} +
\lambda^{(-)}_{m-n,0}) \cr
\phi_n &=& -\pi T\sum_m
\frac{\phi_m}{(|Z_m\omega_m|^2+\phi_m^2)^{1/2}}
(\lambda^{(+)}_{m-n,l}-\lambda^{(-)}_{m-n,l})
\label{eliash}
\end{eqnarray}
where the coupling constants $\lambda$ are obtained by averaging the
effective interactions (\ref{veff}) over the Fermi surface:
\begin{equation}
\lambda^{(\pm)}_{m-n,l} =
\frac{k_f}{2\pi m^*}\int_0^{2k_F} 
\cos\left( 2l\,{\sin^{-1}}\frac{q}{2k_F}\right)
D^{(\pm)}(q,|\omega_m-\omega_n|)
\sqrt{1-(q/2k_F)^2}dq.
\label{cc}
\end{equation}
The pairing interaction mediated by $a^{(-)}$ is singular at small $q$
and is thus attractive in all angular momentum channels.  Here we
consider the case of $s$-wave pairing and henceforth set
$\lambda_{m-n}^{(\pm)} \equiv \lambda_{m-n,0}^{(\pm)}$.  Presumably
residual interactions not considered here will determine which pairing
channel dominates in the actual system.

Performing the integral (\ref{cc}) for $\lambda^{(-)}$ yields
\begin{equation}
\lambda^{(-)}_{m-n} \sim 
\frac{E_f}{(e^2/\epsilon l_0)} 
\left(\frac{l_0}{d}\right)^{2/3}
\left(\frac{e^2/\epsilon l_0}{|\omega_m-\omega_n|}\right)^{1/3} 
+ {\rm\ \  less\ singular\ terms},
\end{equation}
where $E_f = k_f^2/2m^*$ and $l_0 = 1/\sqrt{e B}$ is the
magnetic length, (for $\nu=1/2$, $k_f = l_0^{-1}$). As discussed by
HLR \cite{hlr}, the electron band mass must be renormalized so that
$m^*\sim \epsilon/{e^2} {l_0}$.  For simplicity in what follows we
will take $E_f \sim e^2/\epsilon l_0$ and
\begin{equation}
\lambda^{(-)}_{m-n} = \gamma 
\left(\frac{\omega_0}{|\omega_m - \omega_n|}\right)^{1/3}
\end{equation}
where $\gamma = (l_0/d)^{2/3}$ is a dimensionless `coupling constant',
and $\omega_0 = e^2/\epsilon l_0$.
Performing the same integration for $\lambda^{(+)}$ we obtain
\begin{equation}
\lambda_{m-n}^{(+)} \sim \ln\left(\frac{\omega_0}
{|\omega_n - \omega_m|}\right) + {\rm\ \  less\ singular\ terms}.
\end{equation}

The two Eliashberg equations can be combined to obtain a single
equation for $\Delta_n$:
\begin{eqnarray}
\Delta_n = 
\pi T\sum_m \frac{1}{(\omega_m^2+\Delta_m^2)^{1/2}}
\left[
\left(
{\Delta_m \omega_n-\Delta_n \omega_m \over \omega_n}
\right)
\lambda_{m-n}^{(-)}
-
\left(
{\Delta_m \omega_n+\Delta_n \omega_m \over \omega_n}
\right)\lambda_{m-n}^{(+)}
\right].
\label{eliash2}
\end{eqnarray}
Note that there is a cancellation when $\omega_n = \omega_m$ which
removes the divergence in the attractive interaction
$\lambda_{m-n}^{(-)}$ when $m = n$.  This cancellation can be
understood as a consequence of Anderson's theorem \cite{pwa}. The
quasistatic ($\omega < T$) gauge fluctuations which are responsible
for the destruction of Fermi liquid behavior in the `normal state',
drop out of the gap equation because they can be viewed effectively as
a random time-reversal invariant potential.  Here by time-reversal we
mean combined time-reversal and exchange of the two layers, under
which $a^{(-)}$ is, indeed, invariant. $a^{(+)}$, on the other hand,
is not time-reversal invariant, and, hence is not governed by
Anderson's theorem. As a result, there is a finite-temperature
divergence at $m = n$. The origin of this divergence can be traced
back to the fact that the composite fermion pairs are not `neutral'
with respect to the $a^{(+)}$ field. As a result, the pairing equation
is not gauge-invariant and may contain unphysical divergences.  These
divergences are not present in gauge-invariant quantities such as the
free energy, which was calculated by Ubbens and Lee \cite{ubbens} in a
related problem arising in the gauge theory description of the spin
gap in the cuprates.

We first ignore the $a^{(+)}$ fluctuations and calculate what $T_c$
would be in their absence.  Linearizing (\ref{eliash2}) and setting
$\lambda^{(+)}_{m-n}$ to zero we obtain an equation which, because of
the scaling behavior of $\lambda_{m-n}^{(-)}$, allows the dependence
on the temperature, $\omega_0$ and $\gamma$ all to be factored out.
The resulting equation is
\begin{equation}
\Delta_n =
\gamma\left(\frac{\omega_0}{2\pi T}\right)^{1/3}
\sum_m |m-n|^{-1/3}(1-\delta_{m,n})\left(
\frac{\Delta_m}{2m+1}-
\frac{\Delta_n}{2n+1}\right){\rm sgn}(2m+1).\label{l3}
\end{equation}
Unlike conventional BCS theory there is no need for a frequency cutoff
in the gap equation.  Because the effective interaction falls off as
$\omega^{-1/3}$ it is possible to take the Matsubara sum to infinity.
The resulting expression for $T_c$ in this limit is
\begin{equation}
T_c \simeq 4.3 \omega_0\gamma^3 \propto \frac{1}{d^2}
\end{equation}
where only the proportionality constant needs to be determined
numerically.

We find a similar result for the zero-temperature gap if we continue
to neglect $a^{(+)}$.  The zero-temperature Eliashberg equation on the
imaginary frequency axis can be written
\begin{equation}
\Delta(i\omega) = \frac{1}{2}\int_{-\infty}^{\infty}d\omega^\prime
\frac{1}{({\omega^\prime}^2+\Delta^2)^{1/2}}
\left(
\frac{\omega\Delta(i\omega^\prime)-\omega^\prime\Delta(i\omega)}{\omega}
\right)
\lambda^{(-)}(i\omega-i\omega^\prime).
\label{eliash3}
\end{equation}
Within the approximation $\Delta(i\omega) = {\rm Const}$ the equation
for $\Delta(0)$ obtained by taking the $\omega\rightarrow 0$ limit of
the rhs of (\ref{eliash3}) is
\begin{eqnarray}
1\, &=&\, 
\frac{\gamma{\omega_0}^{1/3}}{3}
\int_{-\infty}^{\infty}d\omega^\prime
\frac{1}{({\omega^\prime}^2+\Delta^2)^{1/2}}
\frac{1}{{\omega^\prime}^{1/3}} \propto \gamma 
\left(\frac{\omega_0}{\Delta}\right)^{1/3}.
\end{eqnarray}
It follows that $\Delta(0) \propto \gamma^3\omega_0$.  A fully
self-consistent solution of (\ref{eliash3}) yields
\begin{equation}
\Delta(0) \simeq 8.4 \gamma^3 \omega_0 \propto \frac{1}{d^2}.
\end{equation}
Thus, in the absence of $a^{(+)}$ fluctuations, the superconducting
energy gap at zero temperature falls off as $1/d^2$.  We emphasize
that the gap, $\Delta \sim \gamma^3
\theta(\gamma)$, is not analytic at $\gamma=0$ and is not a
perturbative effect.

At zero-temperature, the $a^{(+)}$ fluctuations do not lead to any
divergences, so the Eliashberg equations may be solved without special
precaution.  Again, we consider the approximation $\Delta(i\omega) =
{\rm Const}$.  The equation for $\Delta(0)$ is then, in the limit
$\gamma \ll 1$,
\begin{eqnarray}
1\, &=&\,
\frac{\gamma{\omega_0}^{1/3}}{3}
\int_{-\infty}^{\infty}d\omega^\prime
\frac{1}{({\omega^\prime}^2+\Delta^2)^{1/2}}
\frac{1}{{\omega^\prime}^{1/3}}\, - \,
\int_{-\Lambda}^{\Lambda}d\omega^\prime
\frac{1}{({\omega^\prime}^2+\Delta^2)^{1/2}}
\ln\frac{\omega_0}{\omega^\prime}
\cr
&&\cr &\simeq &\, A \gamma \left(\frac{\omega_0}{\Delta}\right)^{1/3}
\,- B \left(\ln\frac{\omega_0}{\Delta}\right)^2 + 
{\rm\ \ less\ singular\ terms.}
\end{eqnarray}
Here $A$ and $B$ are numbers of order 1 and $\Lambda$ is a high-energy
cutoff, $\Lambda\sim{\omega_0}$.  The presence of the $a^{(+)}$
fluctuations lead to a substantial suppression of the gap.  In the
limit $\gamma \ll 1$ we find that
\begin{eqnarray}
\Delta \sim {\omega_0}\frac{\gamma^3}{(\ln\gamma)^6} 
\propto \frac{1}{d^2 (\ln d)^6}.
\end{eqnarray}
Although the gap is suppressed, the $a^{(+)}$ fluctuations do not
eliminate the zero temperature pairing instability.  This is the
central result of this paper --- an ideal DLCFM, as defined above, is
{\it always} unstable to the formation of a paired state no matter how
large the layer spacing is.  

We now comment on the solution of the finite-temperature Eliashberg
equations including the $a^{(+)}$ fluctuations.  As stated above, the
problem is that the logarithmic singularity in $\lambda^{(+)}_{m-n}$
is not canceled by Anderson's theorem.  However, in the presence of a
superconducting gap the $a^{(+)}$ fluctuations also acquire a gap
which in turn cuts off the logarithmic divergence in $\lambda^{(+)}$
when $m=n$ so that $\lambda^{(+)}_{0} \sim \ln(\omega_0/\Delta)$.
Therefore, when treated self-consistently, the $a^{(+)}$ fluctuations
do not lead to any divergence in the finite temperature Eliashberg
equations.  We believe that this treatment is equivalent to the free
energy analysis of Ubbens and Lee \cite{ubbens} and, like them, we
find that within Eliashberg theory the singular $a^{(+)}$ fluctuations
drive the finite-temperature pairing transition first-order.  However,
we do not expect this result to be physically relevant in our case.
Fluctuations about the mean-field Eliashberg treatment presented here
will drive the transition temperature to zero because, as in the
Chern-Simons Landau-Ginzburg theory of the FQHE \cite{zhang}, the
gauge fields screen vortices, rendering their energy finite, rather
than logarithmically divergent.  Thus there is no Kosterlitz-Thouless
transition.

To conclude, we have shown that in the absence of disorder and
interlayer tunneling a perfectly balanced DLCFM, or any system
described by the Lagrangian (\ref{lagrangian}), is {\it always}
unstable to the formation of a paired state at zero temperature,
regardless of how large the layer spacing is.  Such a paired state
will be incompressible and thus exhibit the FQHE \cite{zhang}.
Motivated by this result we propose the qualitative phase diagram
shown in Fig.\ (1) for the $\nu=1$ double-layer system.  The
experimental observation of the paired quantum Hall state discussed in
this paper would provide strong evidence for the existence of gauge
fluctuations in composite fermion metals.

The authors acknowledge useful discussions with B.I. Halperin.  NEB
also acknowledges useful discussions with E. Miranda and J.R.
Schrieffer as well as support from the National High Magnetic Field
Laboratory at Florida State University.  CN would like to thank the
Harvard Physics Department for its hospitality during the completion
of part of this work, and the Fannie and John Hertz Foundation for
support.

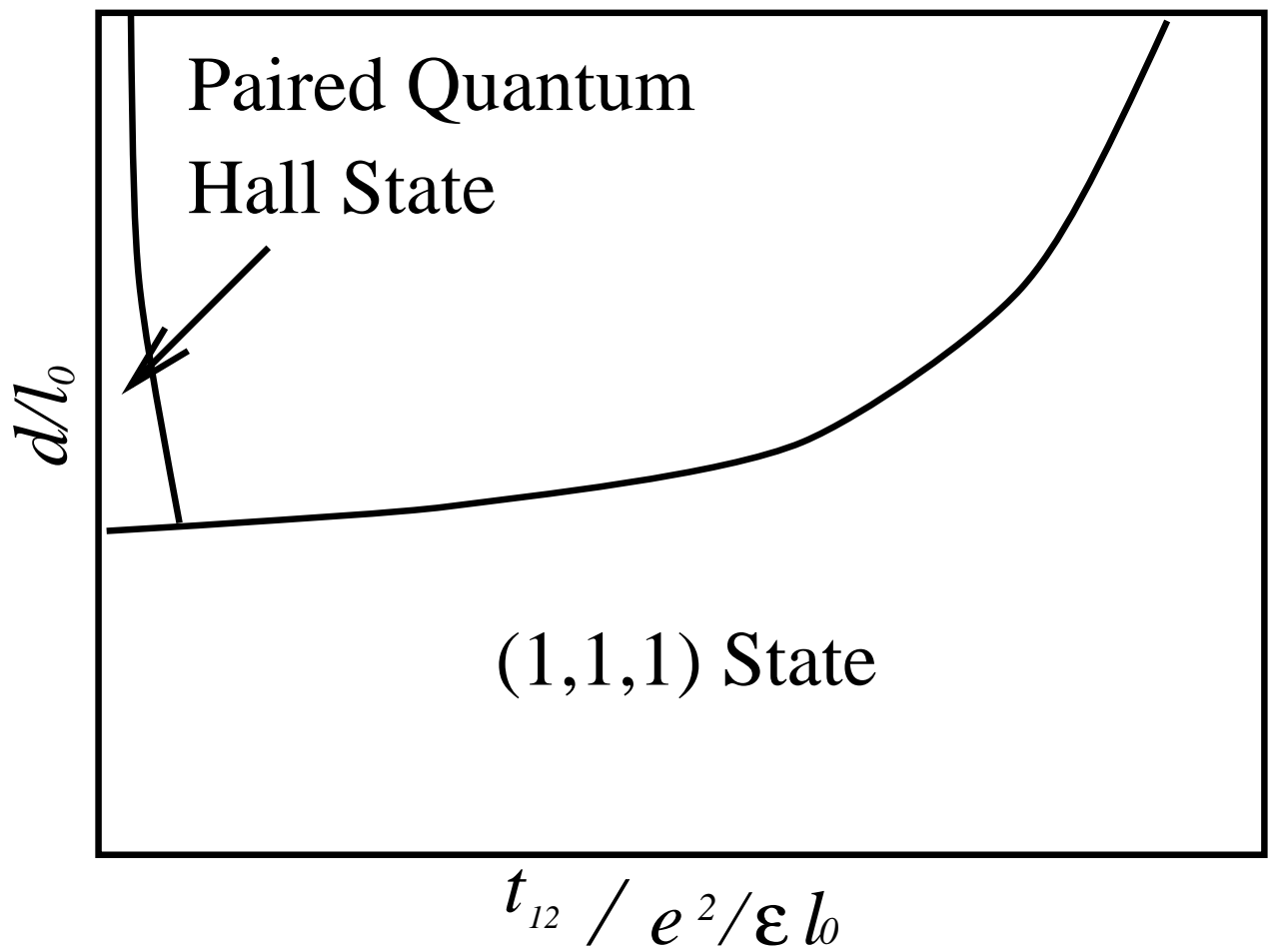
\begin{figure}
\caption{Proposed phase diagram for a double layer $\nu=1$ system including 
the paired quantum Hall state discussed in this paper.  Here $t_{12}$
is the interlayer tunneling amplitude.}

\end{figure} \end{document}